\documentstyle[amssymb,preprint,aps,epsf]{revtex}

\newcommand{\be}{\begin{equation}}
\newcommand{\ee}{\end{equation}}
\newcommand{\bea}{\begin{eqnarray}}
\newcommand{\eea}{\end{eqnarray}}

\begin{document}

\title{ Pathology of Schwinger boson mean field theory for Heisenberg spin models}

\vskip0.5truecm 
\author{Theja N. De Silva, Michael Ma, Fu Chun Zhang }

\address{Department of Physics, University of Cincinnati, OH 45221-0011}
\vskip0.5truecm

\maketitle

\begin{abstract}

{We re-analyze the Schwinger boson mean field theory (SBMFT) for Heisenberg spin models on the cubic lattice. We find that the second
order phase transition point for magnetic ordering previously reported corresponds to a local maximum of the free energy functional.
 For both ferromagnetic and antiferromagnetic Heisenberg
models with spin $S \geq S_{C}$, where $S_{C} < 1/2$, the mean field transitions are first order from the magnetically long-ranged ordered phase to the completely uncorrelated phase. In addition to erroneously giving a first order transition for magnetic ordering, the mean field theory does not include a phase with finite short-range correlation, thus negating one of the prime advantages of SBMFT. The relevance of these pathologies to other situations beyond the cubic lattice is discussed.}

\end{abstract}
\vskip .1truein
\vskip2pc

\newpage
\section{Introduction}
Schwinger boson representation 
is by now a well known tool in the study of quantum spin
systems~\cite{arovas,sarker,yoshioka}. In 
this method, each spin operator is represented by two Schwinger 
bosons. In contrast to Holstein-Primakoff bosons~\cite{holstein}, the
Schwinger boson 
representation of spin operators does not involve square roots. 
Furthermore, while the vacuum state of Holstein-Primakoff bosons has 
to be a broken symmetry state, this is not the case for Schwinger 
bosons. Instead, long-ranged order (LRO) is a result of condensation
of 
the Schwinger bosons~\cite{sarker,yoshioka}. Thus, using Schwinger bosons
allows one to 
naturally address both the magnetically disordered and ordered states.

Using the Schwinger boson representation, the quantum spin 
Hamiltonian is mapped onto an interacting boson 
Hamiltonian, which can then be approximated by a self-consistent 
non-interacting Hamiltonian using mean field theory (MFT). In such a mean 
field theory, a non-zero mean field amplitude corresponds to 
short-ranged order (SRO). Long range order exists only if for a given 
temperature, the mean field Hamiltonian gives rise to Bose 
condensation. The ability of the mean field theory to address both 
short-ranged and long-ranged order is perhaps its most appealing 
feature, especially for 1D and 2D systems, where exact solutions has 
shown that LRO is destroyed at any finite temperature. For the 
Heisenberg ferromagnetic or antiferromagnetic Hamiltonian on a 
d-dimensional hypercubic lattice, the results reported in literature 
are briefly summarized as follows. In both 1D and 2D, in agreement 
with exact results, there is no LRO at any finite temperature. However, SRO persists until a certain temperature is exceeded~\cite{arovas,sarker,yoshioka}. In 3D, LRO is 
stable at low temperature, and there is a continuous phase transition 
into the SRO state at some critical temperature. Within Schwinger 
boson MFT (SBMFT), the order-disorder transition has the usual mean 
field exponents~\cite{sarker}.

In this paper, we re-examine 
the SBMFT for the Heisenberg spin Hamiltonian on the 3D cubic 
lattice. We find that, provided the spin $S$ is greater than some 
critical value $S_{c}$ $(<1/2),$ the previously reported mean field 
solution that exhibits a continuous phase transition  for magnetic LRO 
is actually an unstable solution corresponding to a local 
{\it{maximum}} of the free energy. Instead, the minimum free 
energy solution shows a first order transition. Since it is well 
established that the Heisenberg Hamiltonian has a continuous transition 
in 3D, our analysis raises serious question about the applicability 
of the SBMFT in 3D. Furthermore, the first order transition is into a 
state with zero mean field amplitude, i.e., without any correlation. 
Thus, the supposed advantage of SBMFT of being able to describe SRO 
does not actually exist. These pathologies of the SBMFT are not 
restricted to 3D, but should appear whenever the LRO can be destroyed 
by tuning some parameter. An example of which is the bilayered 
Heisenberg antiferromagnet at $T=0$ where as the inter-layer coupling 
is increased, SBMFT also shows a first order transition into the 
uncorrelated state~\cite{ng}.

The rest of this paper is 
organized as follows. In Section II and III, we will discuss the 
SBMFT solutions for the Heisenberg ferromagnet and antiferromagnet 
separately, showing that the same pathologies are seen in both cases. 
These results will be summarized in Section IV, where we will also discuss their 
relevance to the bilayered Heisenberg antiferromagnet.

\section{Ferromagnetic Heisenberg Model}

In this section we discuss the SBMFT for the FM Heisenberg model. We first briefly review the Schwinger boson representation and its mean field equations~\cite{arovas,sarker,yoshioka}. We then discuss the mean field solutions
of the model with emphasis on 3D. 

We study a spin-S Heisenberg ferromagnet in a d-dimensional hypercubic lattice
with periodic boundary conditions,

\begin{eqnarray}
H=-J\sum_{<i,j>}S_{i} \cdot S_{j},
\end{eqnarray}
where the exchange constant $J>0$ and the sum is over all the nearest-neighbor pairs on the lattice. In the Schwinger boson representation, a spin operator is represented by two types of bosons,

\begin{eqnarray}
S_{i}^{z}=\frac{1}{2}(a_{i\uparrow}^{\dag}a_{i\uparrow}-a_{i\downarrow}^{\dag}a_{i\downarrow}),
\end{eqnarray}

\begin{eqnarray}
S_{i}^{+}=a_{i\uparrow}^{\dag}a_{i\downarrow}, \,\, S_{i}^{-}=a_{i\downarrow}^{\dag}a_{i\uparrow},
\end{eqnarray}
These bosons should satisfy the constraints in their Fock space as follows,
$\sum_{\sigma}a_{i\sigma}^{\dag}a_{i\sigma}=2S$,
with $2S = 1, \, 2, \, ...$ for physical systems.
By introducing a bond operator,
$B_{ij}^{\dag}=\frac{1}{2}\sum_{\sigma}a_{i\sigma}^{\dag}a_{j\sigma}$,
the SU(2) FM Heisenberg Hamiltonian can be written as,
\begin{eqnarray}
H_{FM-B}=-2J\sum_{<ij>} :B_{ij}^{\dag} B_{ij}:+ \frac{NzJ}{2}S^{2}
 + \sum_{i}\lambda_i(\sum_{\sigma}a_{i\sigma}^{\dag}a_{i\sigma}-2S).
\end{eqnarray}

In the above Eqn.,  $: \, :$ denotes a normal ordering of the bosonic operators,
N is the number of sites in the lattice, and $z \, = 2 \, d$ is the nearest neighbor coordinate. A Lagrangian
multiplier field $\lambda_i$ for the constraint on site $i$ is introduced.  Note that
the FM Heisenberg model
is now transformed to an interacting boson model.

To proceed further, a bond mean field amplitude $B=<B_{ij}^{+}>=<B_{ij}>$, assumed to be real and uniform, is introduced. Neglecting fluctuations, the following non-interacting mean field Hamiltonian is obtained.,

\begin{eqnarray}
H_{MF}=\lambda\sum_{i}
(\sum_{\sigma}a_{i\sigma}^{\dag}a_{i\sigma}-2S)-2J\sum_{ij}
B(B_{ij}^{\dag}+B_{ij})+\frac{NzJ}{2}S^{2}+ NzJB^{2},
\end{eqnarray}
In $H_{MF}$ the local constraint at every
site is replaced by a global one, and a
non-zero value of B indicates {\it{ short-range}} spin-spin correlation.
The mean field Hamiltonian can be diagonalized, and we obtain

\begin{eqnarray}
H_{MF}=\frac{NzJ}{2}S^{2}+NzJB^{2}
-2S{\lambda}N+\sum_{\vec k,\sigma}\omega_{\vec k}a_{\vec k \sigma}^{\dag}a_{\vec k \sigma},
\end{eqnarray}
where the energy dispersion is given by $\omega_{\vec k}=JzB(\epsilon_{\vec k}+\Lambda)$,   with  $\epsilon_{\vec k}=\frac{1}{z}\sum_{\vec{\delta}}(1-e^{i\vec k.\vec{\delta}})$, and $\Lambda=\frac{{\lambda}}{JBz}-1.$

The free energy per site is then given by

\begin{eqnarray}
f=\frac{JS^{2}z}{2}+zJB^{2}-2S\lambda -\frac{2}
{{\beta}N}\sum_{\vec k}\log(\frac{1}{1-e^{-\beta\omega_{\vec k}}}),
\end{eqnarray}
where $\beta=\frac{1}{k_{B}T}$, with $k_{B}$ the Boltzmann constant and $T$
the temperature. The amplitudes of the mean fields $B$ and $\lambda$ are
obtained by minimizing the free energy, which leads to two coupled mean field eqns.,

\begin{eqnarray}
S=\frac{1}{N}\sum_{\vec k}\frac{1}{e^{\tilde{\beta}(\epsilon_{\vec k}+\Lambda )}- 1},
\end{eqnarray}

\begin{eqnarray}
B^2=\frac{B}{N}\sum_{\vec k}\frac{1-\epsilon_{\vec k}}{e^{\tilde{\beta}(\epsilon_{\vec k}+\Lambda)}-1},
\end{eqnarray}
where $\tilde{\beta}=1/k_B \tilde{T}$, and $\tilde T = T/(JzB)$ acts as an effective temperature.
In these eqns., $- \Lambda$ may be viewed as an effective chemical potential.
Since $\epsilon_{\vec k}\geq 0$, we have $\Lambda \geq 0$. Magnetic long range order corresponds to Schwinger boson condensation at $\vec k=0$
~\cite{sarker,yoshioka}. Since bosons do not condense at finite temperature in 1D or 2D, the 1D model and
the 2D model are disordered at any temperature. In three and higher dimensions, one expects a finite critical temperature $T_C$, below which 
SBMFT gives a magnetic ordered state.

To study the possible magnetically ordered phase, the Schwinger boson
condensate density $\rho = < a^{\dagger}_{\vec k = 0,
\sigma} a_{\vec k =0, \sigma} > /N$ is introduced, and the coupled mean
field eqns. are rewritten as,

\begin{eqnarray}
S=\rho+\int\frac{d^{d}
\vec k}{(2\pi)^d}\frac{1}{e^{\tilde{\beta}(\epsilon_{\vec k}+ \Lambda)}-1},
\end{eqnarray}

\begin{eqnarray}
B^2 = B(S-\int\frac{d^{d}\vec k}{(2\pi)^d}
\frac{\epsilon_{\vec k}}{e^{\tilde{\beta}(\epsilon_{\vec k}+ \Lambda)}-1}).
\end{eqnarray}

In the Bose condensed state, $\rho > 0$, $\Lambda =0$, and $B \neq 0$; In the disordered
state, $\rho =0$, and $\Lambda > 0$.  Sarker et al.~\cite{sarker} determined
the critical temperature by requiring $\Lambda =0$ and $\rho =0$. This transition point will be labeled as $``a``$, with the effective temperature $\tilde T_a$. By expanding equations 10 and 11 about $\tilde T_a$, the transition was determined to be continuous with mean field exponents. Upon re-examination of the mean field equations, we have, however, found the transition point $\tilde T_a$ to be physically inaccessible if the spin $S$ is greater than some $S_{C}$. To understand this, we note that these equations are well-behaved in term of $\tilde T$. For example, the quantity $B$ is smooth and monotomically decreasing as $\tilde T$ increases as one would expect. This is shown in fig. 1 for the case of $S=1/2$.  At $\tilde T=0$, $B$ is finite, and $\Lambda =0$. As $\tilde T$ increases, $B$ decreases  monotonously. For $\tilde T > \tilde T_a$,  the state is disordered with $\rho=0$; And at 
$\tilde T <  \tilde T_a$, the state is ordered with $\rho > 0$. On the other hand, the physical temperature $T=JzB\tilde T$, and since $B$ decreases with increasing with $\tilde T$, it is possible for $T$ to decrease as $\tilde T$ increases. This is shown in fig. 2. This would mean for example that $B$ would increase  with increasing $T$, which is physically nonsensical. If the second order transition point $\tilde T_a$  belongs to such an unphysical $\tilde T$, it will not be accessible.
 
In ref.2, it was implicitly assumed that the mean field eqns. 9 and 10
have a two branch structure, so that at high $T$, $B=0$ is the only
solution, but below some ${T^{\prime}}$, a second solution with $B \neq 0$
appears. Instead, mean field
eqns.  actually have three branches of solutions for $B$ or $\tilde T$ as
functions
of $T$. One is the trivial solution (upper branch), $B=0$, or $\tilde T
\rightarrow \infty$ at all $T$. The second branch solutions (lower branch)
exist for all $T < T_t$, with $\tilde T$ monotomically increasing with
$T$. For  $T_g < T < T_t$, there is yet a third branch of solutions
(middle branch), where $\tilde{T}$ goes to $\infty$ at $T = T_g$ and
decreases as
$T$ increases until it merges and terminates with the second branch at $T
= T_t$.

The solutions to the mean field 
equations correspond to extrema of the free energy. For $T<T_{t},$ 
there are more than one solutions, and the one giving the lowest free 
energy should be chosen. From the structure of the $\widetilde{T}$ 
vs. $T$ curve, we can deduce that as $T$ is reduced, the free energy 
functional should behave as follows. For $T>T_{t},$ there is only one 
extremum, which is at $B=0.$ Between $T_{t}$ and $T_{g},$ where there 
are three solutions, there is a local minimum at $B=0,$ a local 
maximum at $B=B_{3}$ (middle branch) and a secondary local minimum at 
$B=B_{2}$ (lower branch), with $B_{2}>B_{3}.$ As $T$ decreases in the 
regime, $B_{3}$ moves towards $0$ and at $T=T_{g},$ the curvature at 
$B=0$ changes sign and becomes a local maximum below $T_{g,}$ leaving 
$B_{2}$ as the only minimum. Given this, at some temperature $T_{b}$ 
between $T_{g}$ and $T_{t},$ the free energy at $B_{2}$ will become 
lower than at $B=0.$ Thus, at $T_{b,}$ $B$ will take a jump from $0$ 
to a finite value of $B_{2}.$ In Figs. 3-5, we show plots of free 
energy $f$ vs. $B$ to illustrate the above. In these plots, the 
$f(B)$ is calculated by solving eq. 10 for $\Lambda $ with $B$ as a 
variable and then evaluating $f$ in eq. 7.

Thus, we see that in general, 
SBMFT gives a first order transition in $B.$ Since $B$ is a measure 
of the short range order, the transition in $B$ is fictitious and 
whether it is first order or continuous is not of any real 
significance. More importantly is where the Bose condensation 
transition occurs relative to this transition. Since in terms of 
$\widetilde{T},$ eq. 10 is just that for Bose-Einstein condensation, 
\ $\widetilde{T}_{a} \sim S^{2/d}$ for $d > 2$, as $S\rightarrow 0,$ and
hence 
so does the physical temperature $T_{a}$. Thus for small $S,$ Bose 
condensation occurs at a temperature where the lowest branch is the 
stable solution, and the continuous transition at $\widetilde{T}_{a}$ 
accessible within SBMFT. As $S$ increases, however, 
$\widetilde{T}_{a}$ continues to increase and for 
$S>S_{c} \simeq 0.15,$ 
$\widetilde{T}_{a}>\widetilde{T}_{b},$ where the solution at the 
lower branch becomes meta-stable.  At even larger $S$ (e.g. $S=1/2$), 
$\widetilde{T}_{a}$ becomes $>\widetilde{T}_{t},$ and the solution 
becomes unstable. Either way, for $S>S_{c},$ as the temperature increases 
from zero, the system undergoes a discontinuous transition at $T_{b}$ 
from the LRO state with finite condensate density to the uncorrelated 
state with $B=0.$ There are two problems with this. The first is that 
result of a first order transition in magnetic long range order for the
Heisenberg model in 3D is  
wrong. The second is that one of the major advantages of SBMFT is 
supposedly its applicability to both LRO and SRO state. But in 
fact, the stable solution is either completely uncorrelated or has LRO, 
just like standard Bethe-Weiss MFT.

\section{Antiferromagnetic Heisenberg Model}

In this section, we discuss the SBMFT for the
AFM Heisenberg model described by the Hamiltonian,

\begin{eqnarray}
H=J\sum_{<i,j>}S_{i} \cdot S_{j},
\end{eqnarray}

where $J>0$, and all other notations are the same as in the previous section
for the FM case. We find that the same pathologies in the SBMFT solution
for FM model are
present also in the AFM model. 

The hypercubic lattice is divided into two sub-lattices
{\itshape A\/} and {\itshape B\/}. In a nearest neighbor pair $<ij>$,
$i{\in}{\itshape A\/}$  and $j{\in}{\itshape B\/}$. As in the  FM model,
 two Schwinger bosons $a_{i\sigma}$ with $\sigma={\pm}1$, at each site $i$, and a bond field
$A_{ij}^{\dag}=\frac{1}{2}\sum_{\sigma}{\sigma}a^{\dag}_{i\sigma}a^{\dag}_{j-\sigma}$ 
are introduced.
$A_{ij}$ is  anti-symmetric with respect to the interchange of $i$ and $j$.
 A spin rotation by $\pi$ about the y-axis is made at all the
sites $j$ on the sub-lattice {\itshape B\/}, so that
$a_{j\uparrow}{\longrightarrow}a_{j\downarrow}$,  $a_{j\downarrow}{\longrightarrow}a_{j\uparrow}$. The bond operator is then transformed as 
$A_{ij}^{\dag}\longrightarrow
A_{ij}^{\dag}=\frac{1}{2}\sum_{\sigma}a^{\dag}_{i\sigma}a^{\dag}_{j\sigma}$. This canonical transformation preserves the constraint
 $\sum_{\sigma}a_{i\sigma}^{\dag}a_{i\sigma}=2S=0,1,2,.....$

In terms of bosonic operators, the AFM Heisenberg model reads,

\begin{eqnarray}
H=-\frac{J}{2}\sum_{<ij>}(A_{ij}^{\dag}A_{ij}-2S^{2}).
\end{eqnarray}

The mean field equations of the AFM model may be derived similarly to the FM model. A global Lagrangian multiplier $\lambda$ to replace the constraint on every site, and a real and uniform
mean field amplitude $<A_{ij}>=<A_{ij}^{\dag}>=A$ are introduced. $A \neq
0$ describes short-range AFM
correlations. After some algebra, the mean field Hamiltonian can be written as:

\begin{eqnarray}
H_{MF}=E_{0}+{\lambda}\sum_{k,\sigma}a_{k\sigma}^{\dag}a_{k\sigma}-
\frac{JAz}{2}\sum_{k,\sigma}{\gamma_{k}}(a_{k\sigma}^{\dag}a_{-k\sigma}^{\dag}+h.c),
\end{eqnarray}

with $\gamma_{\vec k}=\frac{1}{z}\sum_{\delta}e^{i\vec k \cdot \vec {\delta}}=1-\epsilon_{\vec k}$,  and $E_{0}=\frac{1}{2}NzJS^{2}-2{\lambda}NS+JA^{2}Nz$.

 $H_{MF}$ is diagonalized by using the standard Bogoliubov transformation, yielding

\begin{eqnarray}
H_{MF}=E_{0}-{\lambda}N+\sum_{k}\omega_{k}(\alpha_{k}^{\dag}\alpha_{k}+\beta_{k}^{\dag}\beta_{k}+1),
\end{eqnarray}

where $\omega_{\vec k}=(\lambda^{2}-(JzA\gamma_{\vec k})^{2})^{1/2}$.
To find the optimal values of the mean fields, the free
energy is differentiated with respect to $A$ and $\lambda$, and the
SBMFT equations for the AFM model are obtained, 

\begin{eqnarray}
S+\frac{1}{2}=\frac{1}{N}\sum_{k}\frac{\mu}{(\mu^{2}-\gamma_{k}^{2})^{1/2}}
\biggl( \frac{1}{e^{\tilde{\beta}(\mu^{2}-\gamma_{k}^{2})^{1/2}}-1}+\frac{1}{2}
\biggr),
\end{eqnarray}

\begin{eqnarray}
A^2=\frac{A}{N}\sum_{k}\frac{\gamma_{k}^{2}}{(\mu^{2}-\gamma_{k}^{2})^{1/2}}\biggl( \frac{1}{e^{\tilde{\beta}(\mu^{2}-\gamma_{k}^{2})^{1/2}}-1}+\frac{1}{2}
\biggr),
\end{eqnarray}

where we have defined an effective inverse temperature $\tilde{\beta}=JAz{\beta}$ and an effective chemical potential $\mu=\frac{\lambda}{JAz}$.  To describe the possible 
Schwinger boson condensed state, condensation density $\rho$ is introduced
 similar to that in the FM model, and the summations are replaced 
by integrals. The mean field eqns. become,

\begin{eqnarray}
S+ \frac{1}{2}=\rho+ \int\frac{d^{d}\vec k} {(2\pi)^d} \frac{1}
{(\mu^2-\gamma_{\vec k}^{2})^{1/2}}
\biggl(\frac{1}{e^{\tilde{\beta}(\mu^2-\gamma_{k}^{2})^{1/2}}-1}+  \frac{1}{2}
\biggr),
\end{eqnarray}

\begin{eqnarray}
A^2=A(\rho+\int\frac{d^{d}\vec k}{(2\pi)^{d}}\frac{\gamma_{\vec k}^{2}}{(\mu^2-\gamma_{\vec k}^{2})^{1/2}}\biggl( \frac{1}{e^{\tilde{\beta}(\mu^2-\gamma_{\vec k}^{2})^{1/2}}-1}+\frac{1}{2}
\biggr)).
\end{eqnarray}

We now discuss the solutions of these eqns. Note that $\gamma_{\vec k}^{2}{\leq}1$, so we have $\mu{\geq}1$. 
Similar to the FM case, $\rho =0$ and $\mu > 1$ in the disordered
phase, and $\rho > 0$ and $\mu =1$ in the Schwinger boson condensation state,
which corresponds to the Neel ordering phase of the spin system. The results are very similar to those for the 3D FM model.

The solutions for the mean field equations are smooth and monotonic functions of the effective temperature
$\tilde T$, but not of the physical temperature $T$.  In Fig. 6, we show the calculated $\tilde T$ as a function of $T$ from the mean field eqns. for a typical spin $S=1/2$. There are three branches of solutions in $\tilde T$.  The $A=0$ branch, and a middle and a lower branches. The Bose condensation point $``a``$ (where $\mu=1$ and $\rho=0$) is again on the upper branch corresponding to a maximum in the free energy. We have calculated the free energy and found that the stable mean field solution
jumps from a high $T$ phase in the  $A=0$ branch to a low $T$ phase
in the lower branch with a finite condensation density at a temperature $T = T_b < T_a$.  Therefore, the transition for LRO is first order. The qualitative feature
of the free energy as a function of $A$ is similar to those in Fig. 3-5 for the FM case. The critical spin for which $T_a=T_b$ for the AFM model
is found to be $S_C \simeq 0.32$, larger than that in the FM model, but still substantially smaller than $1/2$. For $S > S_C$ the ordering transition within SBMFT is first order and for $S < S_C$, the transition is second order.

\section{Conclusions}

We have re-examined the 
Schwinger boson MFT for quantum spin systems and found that for both 
the FM and AFM Heisenberg Hamiltonian, the mean field equations 
possess in general three branches of solutions in the intermediate 
temperature regime. As a consequence, the system in 3D undergoes a 
first order transition from the ordered state to the completely 
uncorrelated state unless the spin $S$ is very small. In addition to 
erroneously predicting a first order transition in magnetic ordering, 
there is no stable solution with finite SRO without LRO, the 
existence of which is supposed to be one of the advantages of SBMFT 
over Bethe-Weiss MFT or spin wave theory.

The three branch structure of 
the mean field solutions is not restricted to 3D. It is the origin 
of the reported discontinuous jump in the short-range MF amplitude $A$ for
the Heisenberg AFM on
2D square lattice as the temperature is increased~\cite{yoshioka}. For the 2D square lattice, however, the effective 
temperature for Bose condensation $\widetilde{T}_{a}=0,$ which means 
$\widetilde{T}_{a}<\widetilde{T}_{b}$ for all $S.$ As a result, SBMFT is
capable in that case of describing the SRO state at low 
temperature. It should not be assumed though that the pathologies 
associated with SBMFT reported in this paper is only relevant for 
$d>2.$ In fact, it will apply whenever the LRO state is destroyed 
by tuning some parameter if that parameter is sufficiently large. 
One example of this is the bilayered Heisenberg AFM Hamiltonian in 
2D, which received interest a few years ago due to its relevance to 
YBCO~\cite{millis,takigawa,tranquada}. In the bilayered system, the
in-plane AF magnetic LRO is 
destroyed by sufficiently large inter-layer coupling which favors 
having inter-plane neighbors forming
singlets~\cite{ng,millis,matsuda,hida1,hida2,dagotto,sandvik1,sandvik,miyazaki,eder}.
Using SBMFT, Ng
et al~\cite{ng} found for $S>S_{c} \simeq 0.35$, the transition is first 
order from finite condensate density to the state of disconnected 
inter-plane singlets with zero in-plane correlations. This result 
contradicted the hypothesis that the 2D bilayered AFM at $T=0$ is in 
the same universality class as the 3D non-linear sigma model at finite $T$, a 
hypothesis partially confirmed by finite sized numerical calculations 
for $S=1/2$~\cite{sandvik}. One possibility is that SBMFT is qualitatively correct, 
but that the value of $S_{c}$ calculated is too small. Subsequently, 
Gelfand et al~\cite{gelfand} provided further evidence that the transition for arbitrary 
$S$ is continuous and that the SBMFT prediction of first order 
transition is incorrect. However, it was not clear why SBMFT gave such a 
qualitatively incorrect result. From the present work, we can now 
understand that the first order transition reported there is an 
artifact of the same pathologies of the SBMFT reported in this paper.

Finally, while we have focused on SU(2) Heisenberg Hamiltonian in this
paper, the results are a consequence of the structure of the mean field
equations, and therefore will be valid for SU(N) generalization of the
Heisenberg Hamiltonian for which the SBMFT approach above~\cite{arovas}
can be applied.

\section{Acknowledgments}

We acknowledge useful discussions with T. K. Ng and K. K. Ng. This work was supported in part by DOE grant No. DE/FG 03-01ER 45687 and by Chinese Academy of Science.

\begin{figure}
\epsfxsize=5.5cm
\centerline{\epsffile{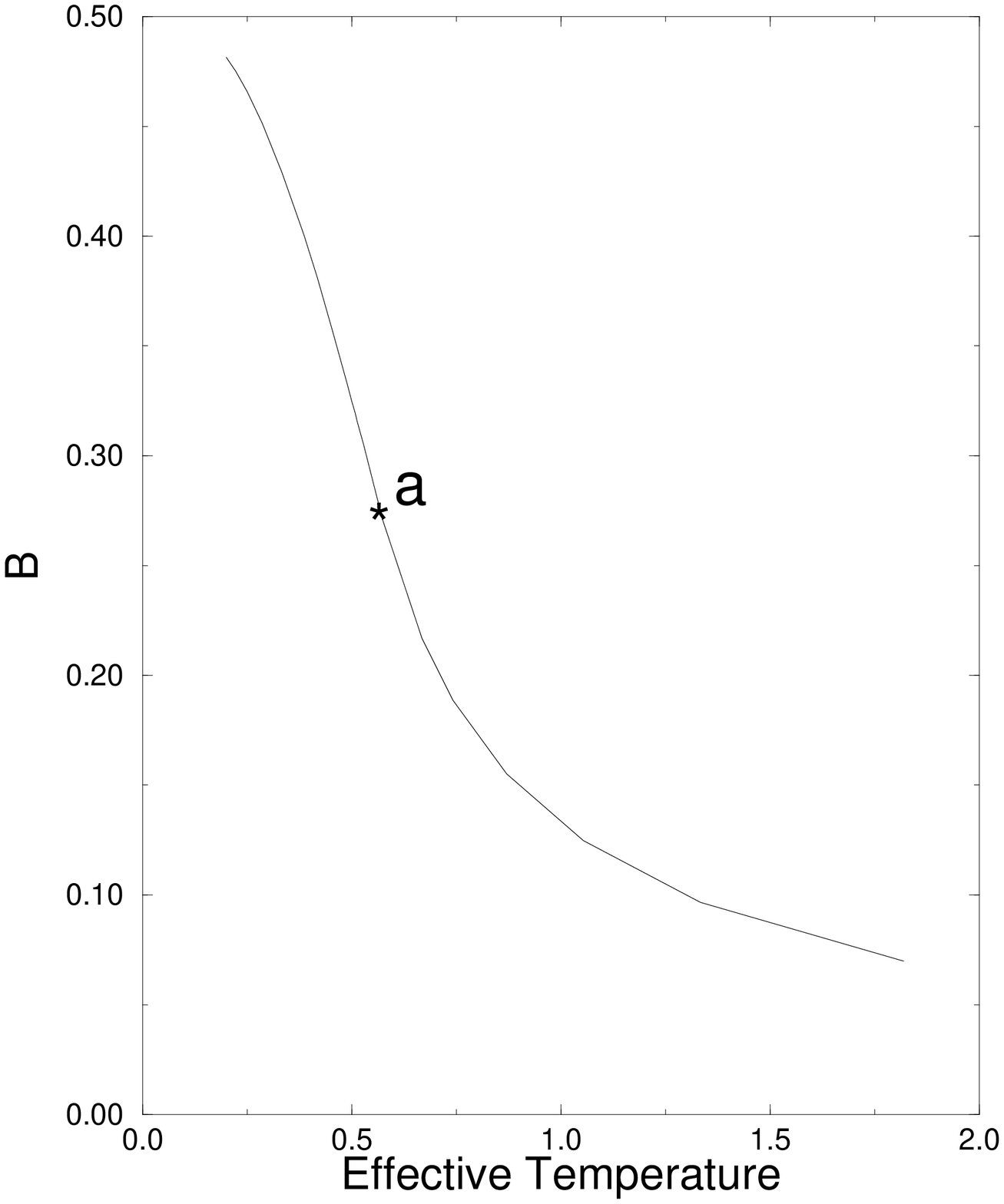}}
\caption{Effective temperature ($\tilde{T}$) dependance of B for three dimensional spin 1/2 Heisenberg ferromagnetic Hamiltonian. Bose condensation occurs at $\tilde{T_{a}}$.}
\label{}
\end{figure}

\begin{figure}
\epsfxsize=5.5cm
\centerline{\epsffile{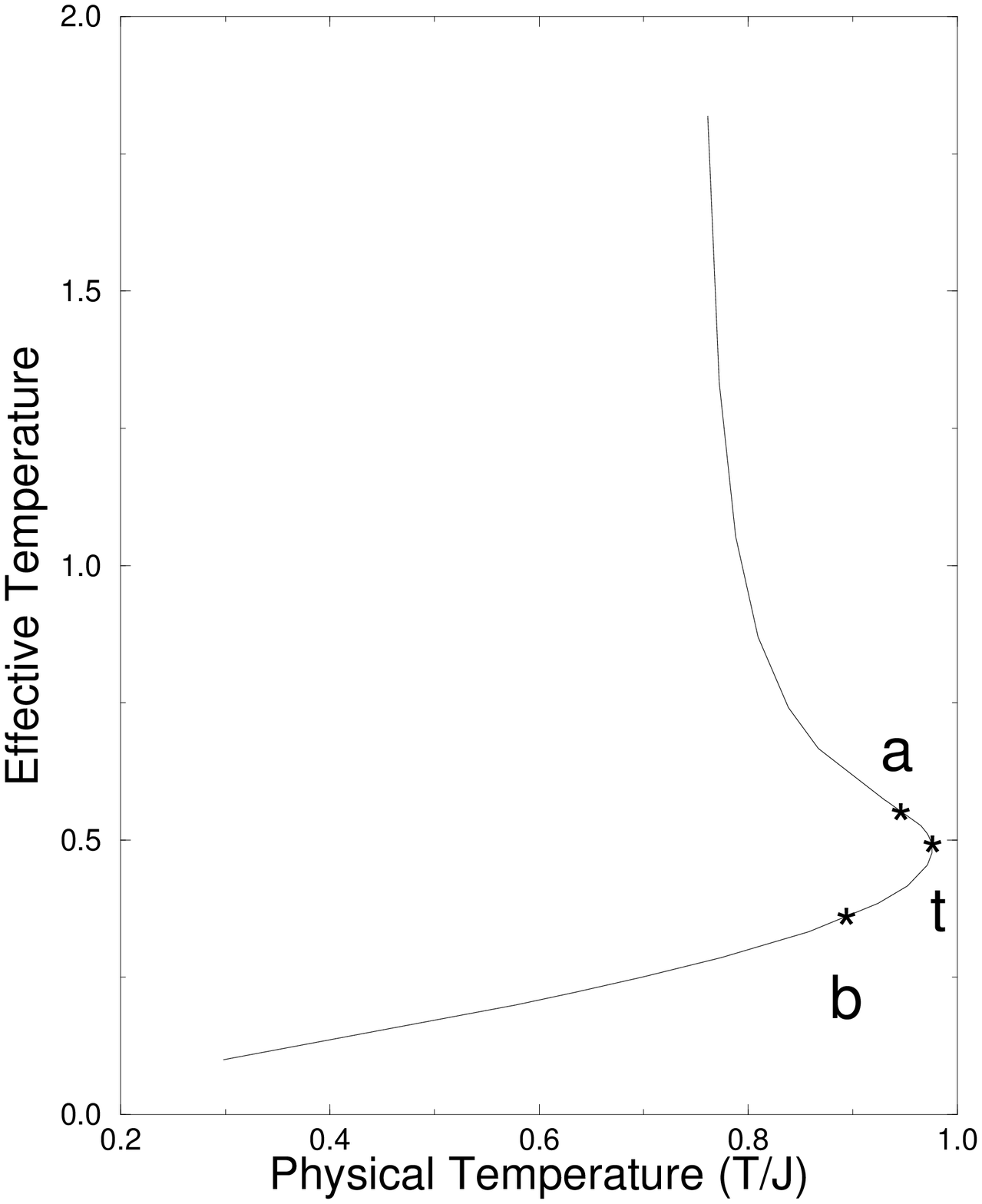}}
\caption{Effective temperature ($\tilde{T}$) as a function of physical temperature $(T/J)$ for three dimensional spin 1/2 Heisenberg ferromagnetic Hamiltonian.}
\label{}
\end{figure}

\begin{figure}
\epsfxsize=5.5cm
\centerline{\epsffile{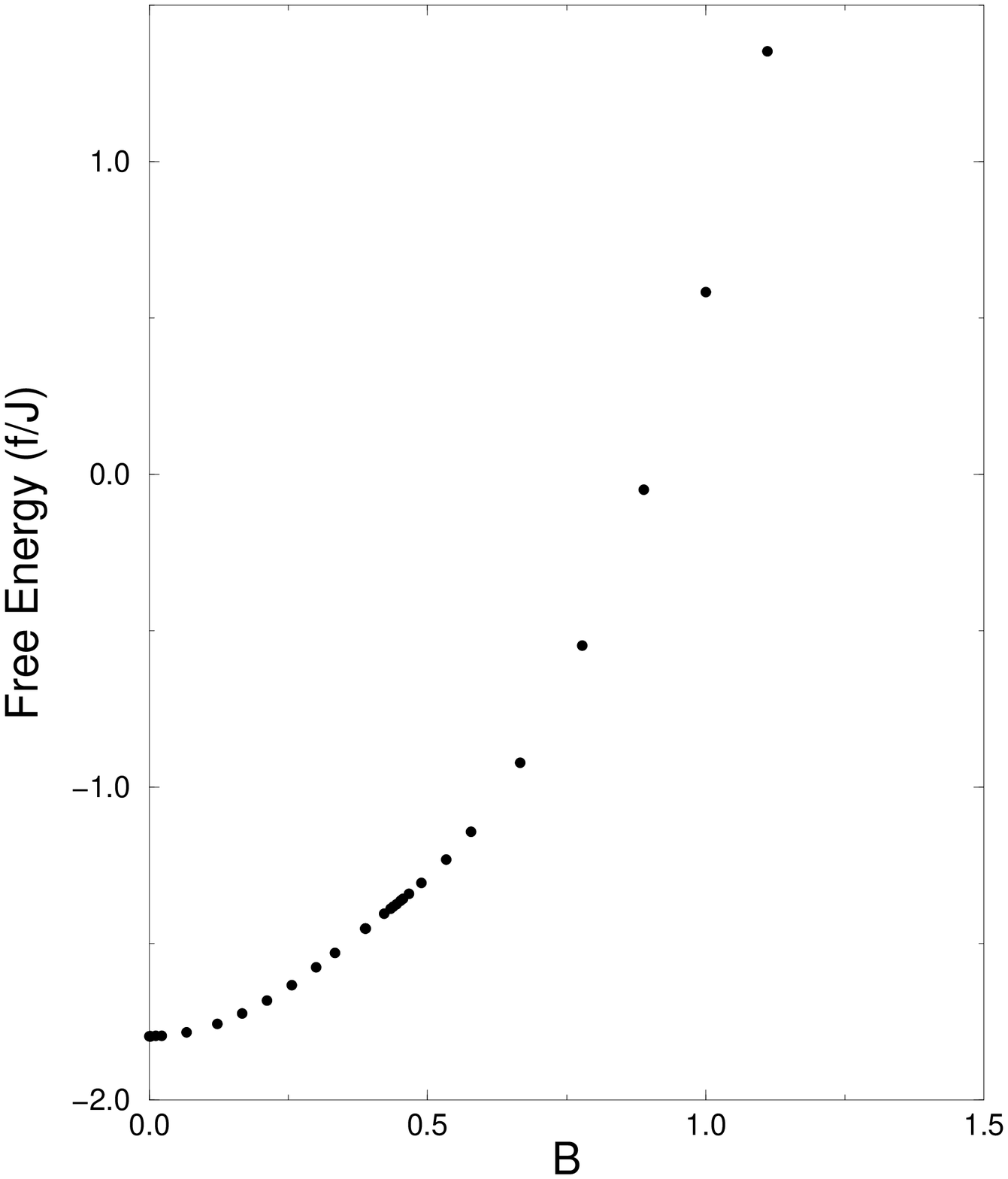}}
\caption{ SB mean field free energy at $T=1.3333J>T_{t}$ for three dimensional spin 1/2 Heisenberg ferromagnetic Hamiltonian.}
\label{}
\end{figure}

\begin{figure}
\epsfxsize=5.5cm
\centerline{\epsffile{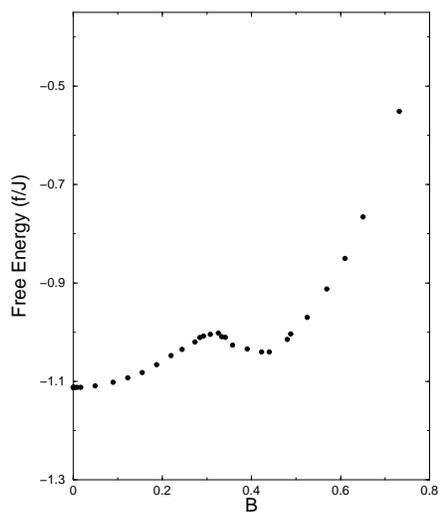}}
\caption{SB mean field free energy at $T_{b}<T=0.9753J<T_{t}$ for three dimensional spin 1/2 Heisenberg ferromagnetic Hamiltonian.}
\label{}
\end{figure}

\begin{figure}
\epsfxsize=5.5cm
\centerline{\epsffile{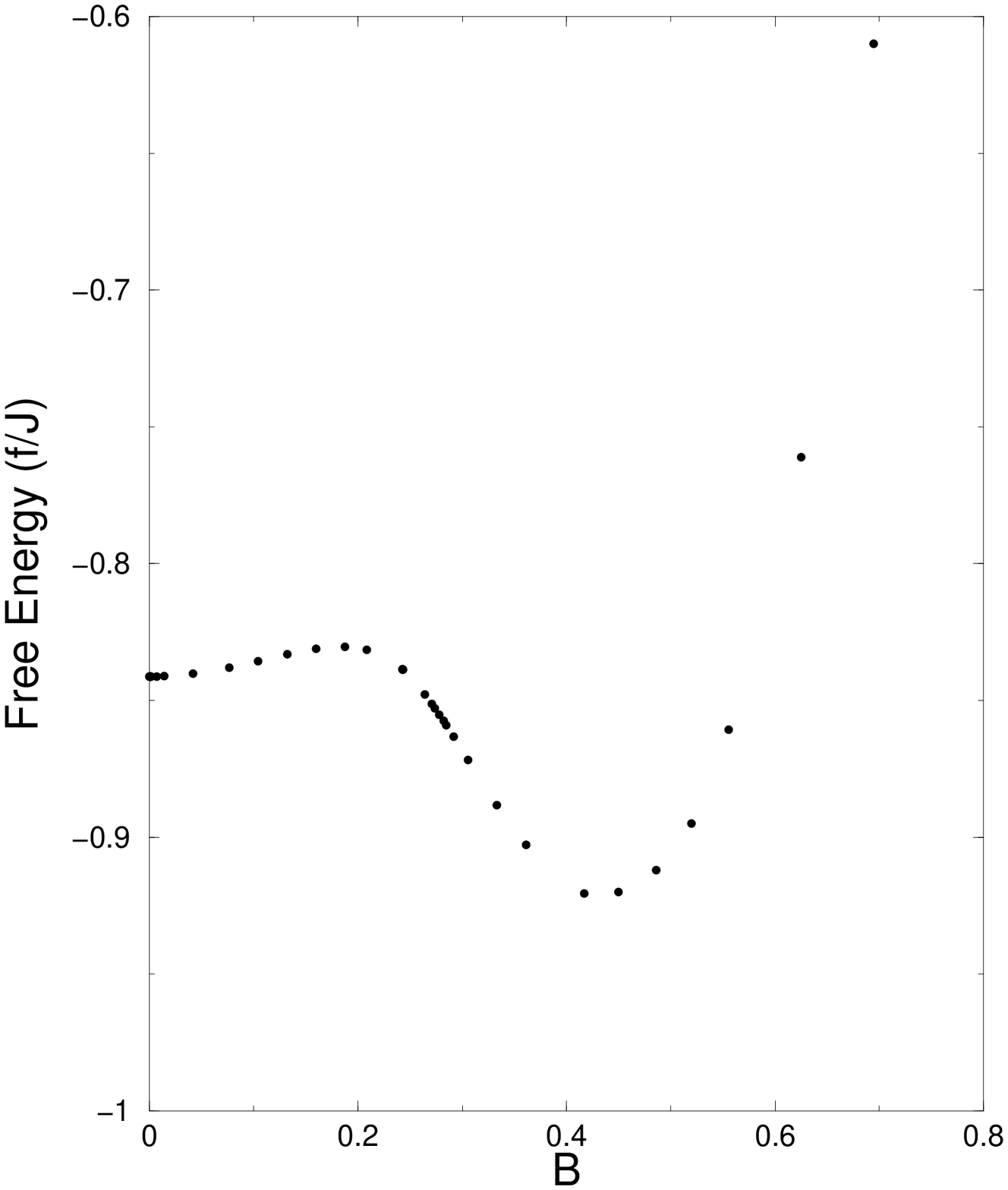}}
\caption{SB mean field free energy at $T=0.8333J<T_{b}$ for three dimensional spin 1/2 Heisenberg ferromagnetic Hamiltonian.}
\label{}
\end{figure}

\begin{figure}
\epsfxsize=5.5cm
\centerline{\epsffile{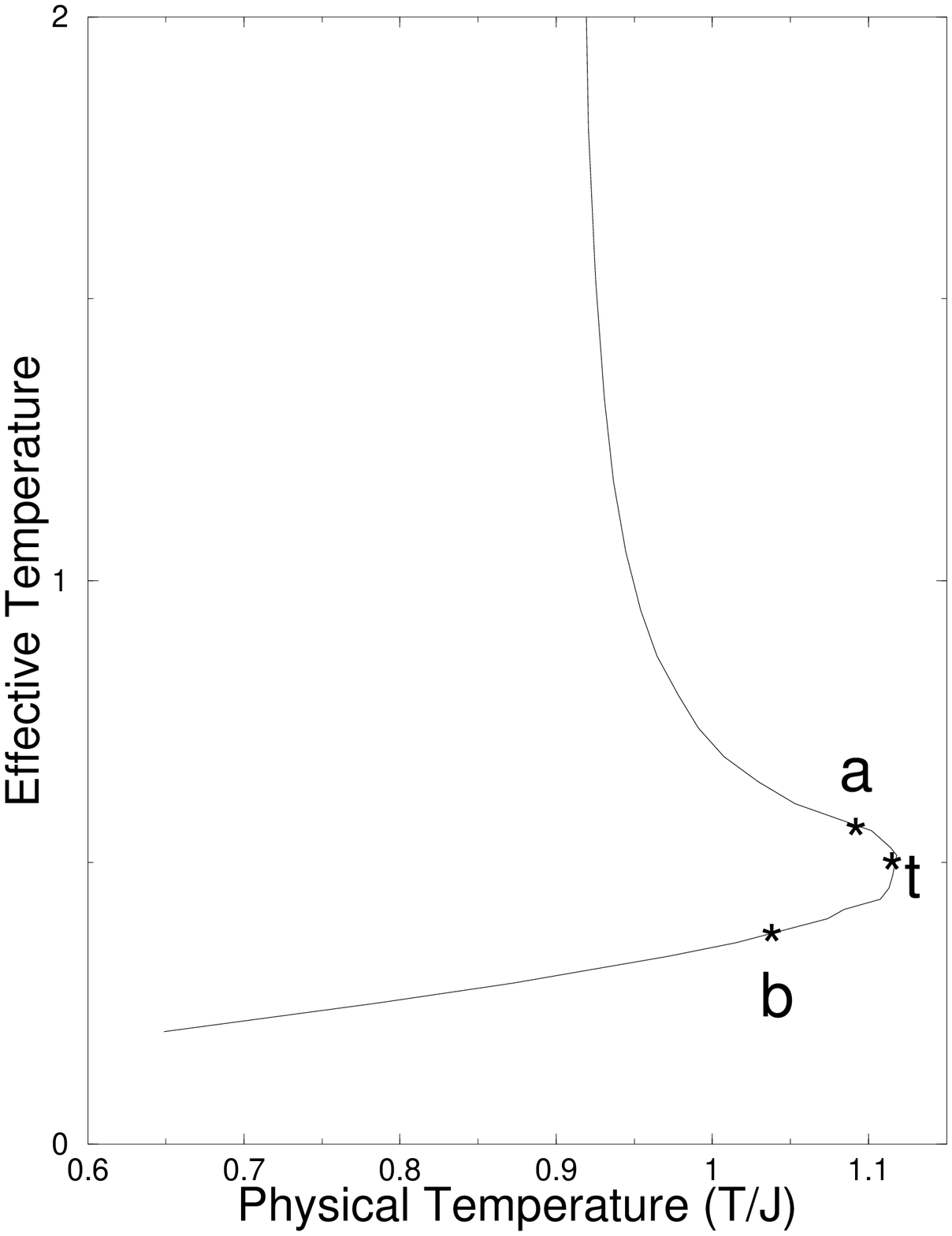}}
\caption{Effective temperature ($\tilde{T}$) as a function of physical temperature $(T/J)$ for three dimensional spin 1/2 Heisenberg antiferromagnetic Hamiltonian.}
\label{}
\end{figure}

\end{document}